\documentclass{osa-article}

\usepackage{amsmath}
\usepackage{amsfonts}
\usepackage{amssymb,upref}
\usepackage{tgtermes} 
\usepackage{anyfontsize}
\usepackage{xcolor}
\usepackage{graphicx}
\usepackage{enumitem}
\usepackage{subfig}
\usepackage{siunitx}
\usepackage[utf8]{inputenc}
\usepackage[T1]{fontenc}

\journal{osac}

\articletype{Research Article}

\begin{document}

\title{A quantum-inspired Fredkin gate based on spatial modes of light}

\author{Daniel F. Urrego,\authormark{1,*} Dorilian Lopez-Mago,\authormark{2} Ver\'onica Vicu\~na-Hern\'andez,\authormark{1} and Juan P. Torres \authormark{1,3}}

\address{\authormark{1}ICFO -- Institut de Ciencies Fotoniques, The Barcelona Institute of Science and Technology, 08860 Castelldefels, Barcelona, Spain\\
\authormark{2}Tecnologico de Monterrey, Escuela de Ingenier\'ia y Ciencias, Ave. Eugenio Garza Sada 2501, Monterrey, N.L. 64849, Mexico\\
\authormark{3}Department of Signal Theory and Communications, Universitat Politecnica de Catalunya, Barcelona, Spain}

\email{\authormark{*}daniel.urrego@icfo.eu} 



\begin{abstract}
  Insights gained from quantum physics can inspire novel classical technologies. These {\em quantum-inspired} technologies are protocols that aim at mimicking particular features of quantum algorithms.  They are generally easier to implement and make use of intense beams. Here we demonstrate in a {\em proof-of-concept} experiment a quantum-inspired protocol based on the idea of quantum fingerprinting (Phys. Rev. Lett. {\bf 87}, 167902, 2001).The carriers of information are optical beams with orbital angular momentum (OAM). These beams allow the implementation of a Fredkin gate or polarization-controlled SWAP operation that exchanges data encoded on beams with different OAM. We measure the degree of similarity between waveforms and strings of bits without unveiling the information content of the data. 
\end{abstract}

\section{Introduction}
The capacity to transmit and process classical and quantum information has experienced tremendous growth in the latest years~\cite{Hilbert}. However the need to continue this trend poses challenges in areas such as computing, nanotechnology, telecommunications, and information processing~\cite{TARAPHDAR}. One promising direction to handling increasingly huge sets of data is to build information-processing devices based on optical logic gates. 

These gates make use of light beams with information encoded in their field amplitude and polarization. At the quantum level they use single photons with the information embedded in their quantum state. A reversible logical gate that has received great attention is the Fredkin gate, or controlled-SWAP (c-SWAP) gate, introduced by Edward Fredkin in the context of computational models to perform any logical or arithmetic operation in the domain of reversible logic-based operations. This gate has three input bits and three output bits and swaps or not the last two bits depending on the value of the first bit that acts as control bit~\cite{Fredkin}. A generalized version of the Fredkin gate allows direct estimations of linear and nonlinear functionals of a quantum state~\cite{Ekert}.

There have been experimental and theoretical proposals to implement a Fredkin gate with optical systems~\cite{Hardy,Cohen}. Additional theoretical work has considered a quantum version using single atoms and single photons~\cite{kumar,Gong,Fuirasek,wang}. Current experimental work includes nuclear magnetic resonance (NMR)~\cite{Du}, superconducting quantum circuits~\cite{Liu}, DNA enzymes~\cite{Orbach}and weak coherent pulses~\cite{Xu,Clarke}. 

Recently there has been the first demonstrations of a quantum Fredkin gate using linear optics with quantum-entangled photons~\cite{Patel,Ono2017,Starek2018}. Generally speaking these implementations are probabilistic and experimentally cumbersome, requiring the use of multiple interferometers. The low efficiency of photonic quantum gates makes the success rate of these gates extremely low. For instance, in ~\cite{Patel} the experimental setup makes use of three interferometers and the successful operation of the Fredkin gate requires the measurement of fourfold coincidences across four single-photon detectors. They measure a fourfold concidences rate of $2.2$ per minute. In~\cite{Ono2017} they use several interferometers that should be perfectly stabilized yielding a low count rate of 10$^{-4}$ Hz. In~\cite{Starek2018} the experimental setup is composed of several Mach–Zehnder interferometers with seven independent phases that makes the whole system prone to imperfections.

One interesting and promising option that attracts a lot of attention is the implementation of classical logical gates whose design is inspired by counterpart quantum gates \cite{Baris,Banaszek,Kaltenbaek,Kaltenbaek2008,Lavoie,Ogawa,Toninelli,Qian,Kagalwala,Mazurek,Altmann}. They are generally not probabilistic, much easier to implement and make use of intense beams. The transformation of protocols and technologies designed with quantum tools into classical protocols and technologies is based on the fact that certain features of quantum physics are also shared by waves in the classical world. This is the case of interference or entanglement between degrees of freedom of a single particle. However there will aspects of these quantum-inspired gates, such as non-locality, that will make fundamentally different classical analogs from its counterpart quantum algorithms. How far one can go in this analogy is a matter of discussion and controversy in the science community~\cite{Markiewicz,Qian}.

Here we demonstrate in a {\em proof-of-concept} experiment a quantum-inspired protocol for comparing strings of data and waveforms without the need to unveil the information contained in the signals. For implementing this gate,  we translate key ideas and elements of the quantum fingerprinting protocol \cite{Buhrman} to the classical domain. One key element is a Fredkin gate that mimicks features of the counterpart quantum-optical Fredkin gate originally proposed by Milburn~\cite{Milburn}. The two channels or carriers of information are two strings of orthogonal spatial modes with orbital angular momentum (OAM). Information is encoded in various formats on these two channels.  OAM beams allow implementing a crucial element of this system: a polarization-controlled SWAP operation. The c-SWAP operation exchanges or not data carried by different channels depending on the state of polarization of the beam, that acts as control bit. Our system can compare two strings of data and evaluate the degree of similarity of the information encoded in the complex amplitudes. We will show below several examples of this that will validate the capability of our system for estimating the fidelity of streams of data without evaluating the data itself.

\begin{figure}[t!]
\centering
\includegraphics[width=6cm]{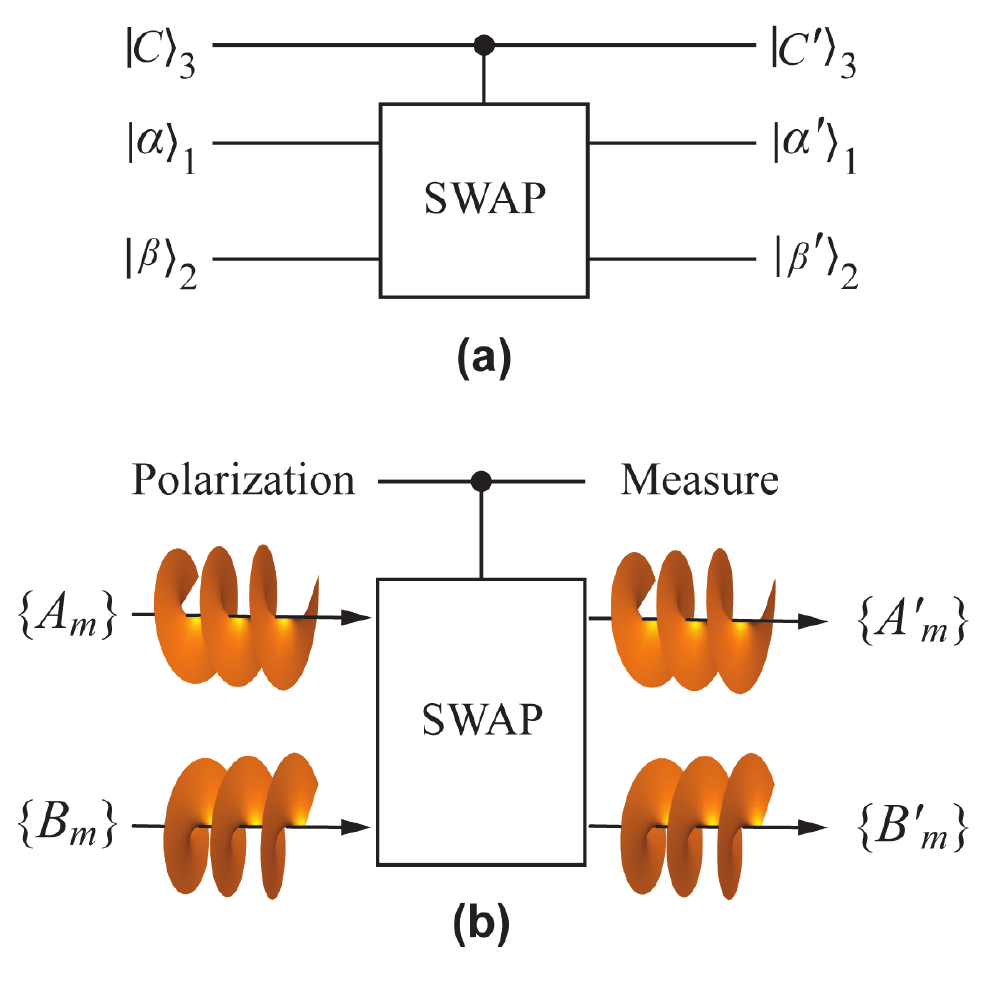}
\caption{The Fredkin gate. (a) Quantum Fredkin gate. The quantum state of input channels $1$ and $2$ is either swapped or not depending on the value of the control bit $|C\rangle_3$. (b) Quantum-inspired controlled-swap gate built with spatial modes carrying orbital angular momentum. Indexes $m$ and $-m$ swap their sign or not depending on the polarization of the beam. The state of polarization determines if the gate perform the identity transformation ($\{ A_m^{\prime} \}=\{ A_m \}$ and $\{ B_m^{\prime} \}=\{ B_m \}$) or the swap transformation ($\{ A_m^{\prime} \}=\{ B_m \}$ and $\{ B_m^{\prime} \}=\{ A_m \}$).}   
\label{fig-gate-depectation}
\end{figure} 

Spatial modes of light play a central role in the development of new information technologies, information processing, and secure communications. OAM modes are particularly interesting for quantum and classical communications due to its capacity for carrying large amounts of information~\cite{Baccon}. Interferometric methods can be used for measuring and sorting modes with OAM in the classical and quantum domains~\cite{Leach2002}. When considering polarization-sensitive interferometers with Dove prims capable of performing NOT operations on the sign of OAM modes, one can generate and manipulate OAM beams with high efficiency and robustness~\cite{Slussarenko2010}. In the last few years a lot of attention has been directed towards using such beams for information transfer in the context of free-space communications~\cite{Jian,Djordjevic}. 

\section{The Fredkin gate in the quantum and classical domains}
The circuit representation of the original quantum Fredkin gate is shown in Fig. ~\ref{fig-gate-depectation}(a). It is a 3-qubit gate that performs a c-SWAP operation conditioned by the state of the control qubit $|C\rangle_3$. At the input we have qubit $|\alpha \rangle_1$ in channel 1 and qubit $|\beta \rangle_2$ in channel 2. If the control bit is $|0\rangle_3$, the qubits in each channel remain the same: $|\alpha^{\prime} \rangle_1=|\alpha \rangle_1$ and $|\beta^{\prime} \rangle_2=|\beta \rangle_2$.  If the control qubit is $|1\rangle_3$, the qubits are swapped between channels: $|\alpha^{\prime} \rangle_1=|\beta \rangle_1$ and $|\beta^{\prime} \rangle_2=|\alpha \rangle_2$.
 
In our quantum-inspired Fredkin gate, the two channels correspond to a set of Laguerre-Gauss spatial modes $\mathrm{LG}^{0}_{m}({\bf r}_{\perp})$ with either positive or negative index $m$. Modes with positive index $m$ correspond to channel 1 and modes with negative $m$ correspond to channel 2. This index indicates a varying phase of the field of the form $\sim \exp(i m \varphi)$, where $\varphi$ is the azimuthal angle in cylindrical coordinates. $m$ also designates an OAM content of $m\hbar$ per photon of the mode. $p=0$ is the radial index of the modes and ${\bf r}_{\perp}=(x,y)$ is the transverse coordinate. 

Information in channel 1 is encoded into $N$ complex amplitudes $A_{m}$ and information in channel 2 is similarly encoded into $N$ complex amplitudes $B_{m}$. Each channel can contain one or several modes, and each amplitude ($A_m$ or $B_m$) can take an array of different values. For the sake of clarity let us consider some examples. Bits can be implemented using channels containing a single mode ($m=1$) with each amplitude $A_1$ and $B_1$ taking one of two values. These two values can be two different phases: $0$ and $\pi$. If each amplitude can take one of three values we would be implementing trits. Four possible values would yield quarts. Another option to implement quarts is to consider channels composed of two modes ($m=1$ and $m=2$) with amplitudes $A_1$, $A_2$, $B_1$ and $B_2$ taking one of two values. In general, the amplitude of the electric field writes
\begin{equation}
E({\bf r}_{\perp}) = \sum^{N}_{m = 1} \left[ A_{m} \mathrm{LG}_{m}^{0}({\bf r}_{\perp}) + B_{m}\mathrm{LG}_{-m}^{0}({\bf r}_{\perp}) \right]. 
\label{equ:superposition}
\end{equation}
The role of control bit in our implementation of the Fredkin gate is the polarization of the spatial modes.

Figure \ref{fig-gate-depectation}(b) shows a schematic representation of the c-SWAP gate between modes with positive and negative index $m$ conditioned by the state of polarization.  For this, we use a Mach-Zehnder interferometer where each arm of the interferometer bears a different orthogonal polarization. In each arm, the beam experience a different number of reflections. A single reflection in a mirror changes the OAM of the LG modes $m \Longleftrightarrow -m$.  In the arm of the interferometer with vertical polarization the beam experience an odd number of reflections that implements the SWAP operation, the electric field amplitude changes as
\begin{eqnarray} 
& & \sum^{N}_{m = 1}  \left[A_{m} \mathrm{LG}_{m}^{0}({\bf r}_{\perp}) + B_{m} \mathrm{LG}_{-m}^{0} ({\bf r}_{\perp}) \right] \Rightarrow  \sum^{N}_{m = 1} 
\left[ B_{m} \mathrm{LG}_{m}^{0}({\bf r}_{\perp}) + A_{m} \mathrm{LG}_{-m}^{0} ({\bf r}_{\perp}) \right].
\label{swap_vertical}
\end{eqnarray}
In the other arm, with horizontal polarization, the beam experiences an even number of reflections, we have the identity transformation. The electric field amplitude changes as
\begin{eqnarray}
 & & \sum^{N}_{m = 1} 
\left[ A_{m} \mathrm{LG}_{m}^{0}({\bf r}_{\perp}) + B_{m} \mathrm{LG}_{-m}^{0} ({\bf r}_{\perp}) \right] \Rightarrow  \sum^{N}_{m = 1} 
\left[ A_{m} \mathrm{LG}_{m}^{0}({\bf r}_{\perp}) + B_{m} \mathrm{LG}_{-m}^{0} ({\bf r}_{\perp}) \right]. 
\label{identity_vertical}
\end{eqnarray}

From an experimental point of view, due to the symmetry of channels $1$ and $2$ with respect to the sign of the index $m$, we can effectively perform the polarization-dependent SWAP operation of amplitudes $\left\{A_m,B_m \right\}$ by performing a polarization-dependent CNOT operation on the modes that compose the channel. Information is contained on the amplitudes, while the sign of index $m$ designates the channel. In an alternative scenario where channels $1$ and $2$ would correspond to single spatial modes with OAM indexes $m_1$ and $m_2$, one can exchange information between channels using an spatial light modulator encoded with $-m_1-m_2$ as demonstrated in~\cite{wang}.

\section{A quantum-inspired optical device for data and waveform comparison}

The system we demonstrate consists of: 1) A Hadamard operation in polarization, the degree of freedom that plays the role of the control bit in our scheme, 2) a Fredkin gate as discussed above, and 3) another Hadamard operation in the polarization degree of freedom. The relevant measurement for data comparison between information encoded in channels $1$ and $2$ is the output power in the horizontal ($P_x$) and vertical ($P_y$) polarizations.

We define the overlap $\gamma$ as
\begin{equation}
\gamma = \dfrac{P_{y} - P_{x}}{P_{y} + P_{x}}.
\label{fredkin-overlap}
\end{equation}
One can easily show that the overlap is related to the values of strings $A_m$ and $B_m$ (see Appendix~\ref{sec: methods}) as
\begin{equation}
\gamma = -\dfrac{ \sum_{m=1}^{N} \left( A_m B_m^*+A_m^* B_m \right)}{\sum_{m=1}^{N} \left(\vert A_{m}\vert^{2}+\vert B_{m} \vert^2 \right)}.
\label{eq: gamma}
\end{equation}
If the two strings of complex numbers are equal ($A_{m} = B_{m}$), we have  $P_{y}=0$ and $\gamma = -1$. If there is a $\pi$ phase difference between them ($A_{m} = -B_{m}$), we have $P_x=0$ and $\gamma=1$. If the two strings are orthogonal, i.e., there is no $m$ for which both $A_m$ and $B_m$ are nonzero, $P_x=P_y$ and $\gamma=0$. In general, the overlap is a real number between $-1$ and $1$. 

In order to unveil the meaning of the parameter $\gamma$, let us assume that $\{A_m\}$ and $\{B_m \}$ are real and that $p_m \equiv |A_m|^2$ and $q_m \equiv |B_m|^2$ ($m=1,2,\ldots,N$) correspond to two probability distributions. We obtain that $\gamma=-\sum_{m=1}^{N}\sqrt{p_m q_m}$. This shows that the overlap measure introduced for a series of complex numbers is related to the fidelity or Bhattacharyya coefficient~\cite{Fuchs}, a measure of how different are two probability distributions.

We consider now that signals $A_m$ and $B_m$ can vary in time. One can think of the discretization of signals of interest at times $t_1=0, t_2=\Delta t, t_3=2\Delta t, \ldots, t_{z}=(Z-1) \Delta t$.  For the case $N=1$ (single-mode) we consider two functions $\alpha_1(t_i)$ and $\beta_1(t_i)$ ($i=1,\ldots,Z$) that correspond to two probability distributions. We encode their values into the phases of $A_1$ and $B_1$, i.e., $A_1(t_i)=\exp[i\alpha_1(t_i)]$ and $B_1(t_i)=\exp[i\beta_1(t_i)]$. We obtain $\gamma(t_i)=-\cos[\alpha_1(t_i)-\beta_1(t_i)]$. The Kolmogorov distance $K(\alpha,\beta)=\sum_{i=1}^{Z} \left| \alpha_1(t_i)-\beta_1(t_i)\right|$ between the two probability distributions is  
\begin{equation}
K(\alpha,\beta)=\sum_{i=1}^{Z}\,\left|\cos^{-1} [\gamma(t_i)] \right|.
\end{equation}

\section{Experimental setup}
\label{experimental-section}

\begin{figure}[t!]
\centering
\includegraphics[width=8cm]{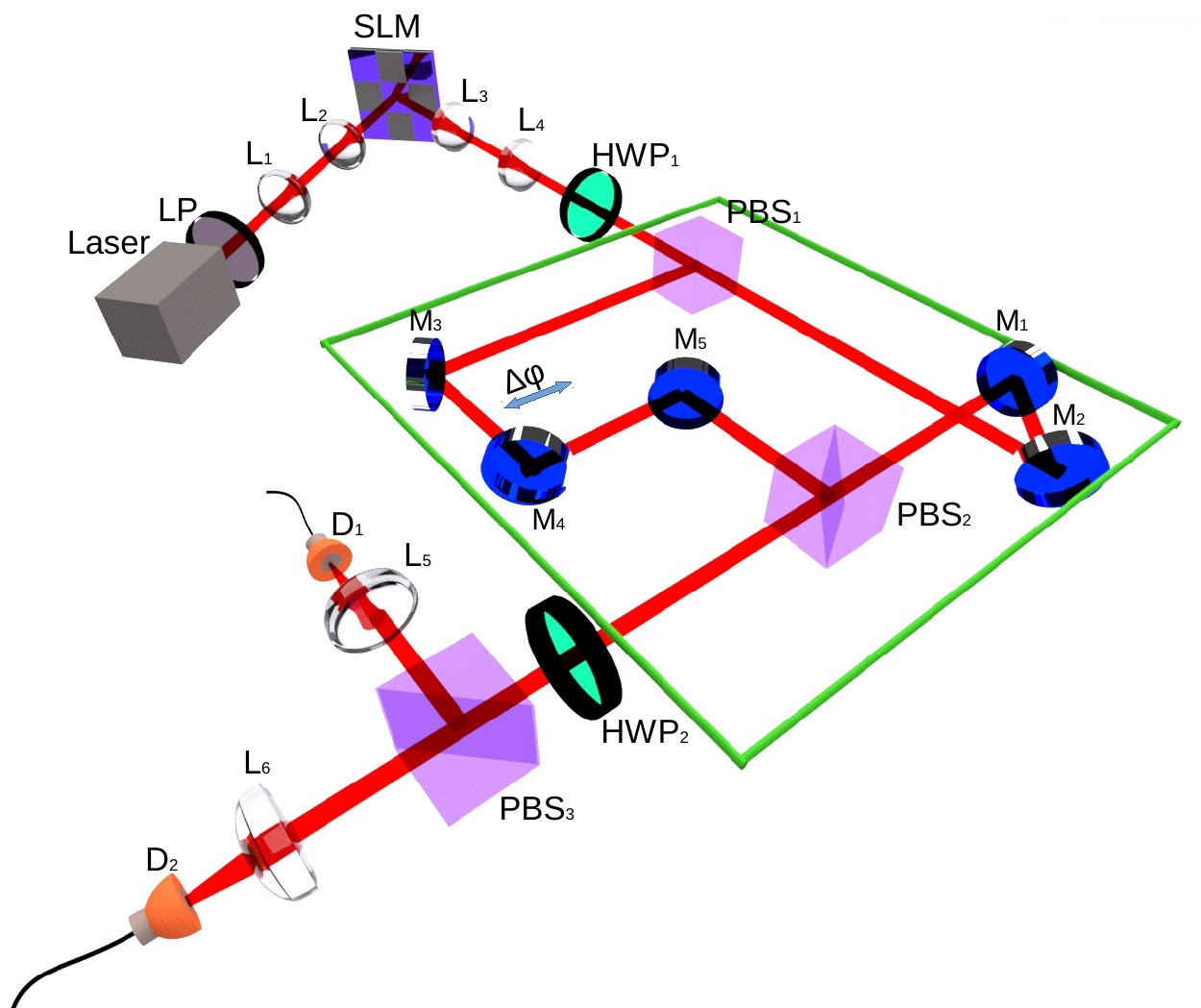}
\caption{Experimental setup. Beams with OAM are generated with the help of a spatial light modulator (SLM). The Hadamard gates are implemented with half-wave plates oriented at $22.5^{\circ}$ with respect to the horizontal polarization. The c-SWAP gate is a Mach-Zehnder interferometer where each arm bears a different polarization. PBS$_i$= polarizing beam splitter; L$_i$: lenses; HWP$_i$: Half-wave plates; M$_i$: mirrors; D$_i$: photodetectors; $\Delta \varphi$: adjustable phase.}
\label{fig:setup}
\end{figure}

Experimental implementation of the protocols for waveform and data comparison that includes the quantum-inspired Fredkin gate, is shown in Fig.~\ref{fig:setup}. We use a Gaussian beam from a Helium-Neon laser ($\lambda=633$~nm) with a beam waist of $\sim$ 1.4~mm. The beam shows vertical polarization with the help of a linear polarizer (LP). It is collimated by two lenses, L1 and L2, with focal lengths of 10~cm and separated by 20~cm. 

We generate superpositions of LG modes with positive and negative OAM indexes ($\pm m$) with the help of a Spatial Light Modulator (SLM, Hamamatsu X10768-01, 792 $\times$ 600 pixels with a pixel pitch of 20 $\mu$m). The spatially-dependent phase of the incoming beam is tailored with appropriate computed-engineered phase patterns displayed on the SLM. A half-wave plate (not shown in the figure) changes the polarization orientation of the beam to horizontal as required by the SLM.

\begin{figure}[t!]
\centering
\includegraphics[width=8.5cm]{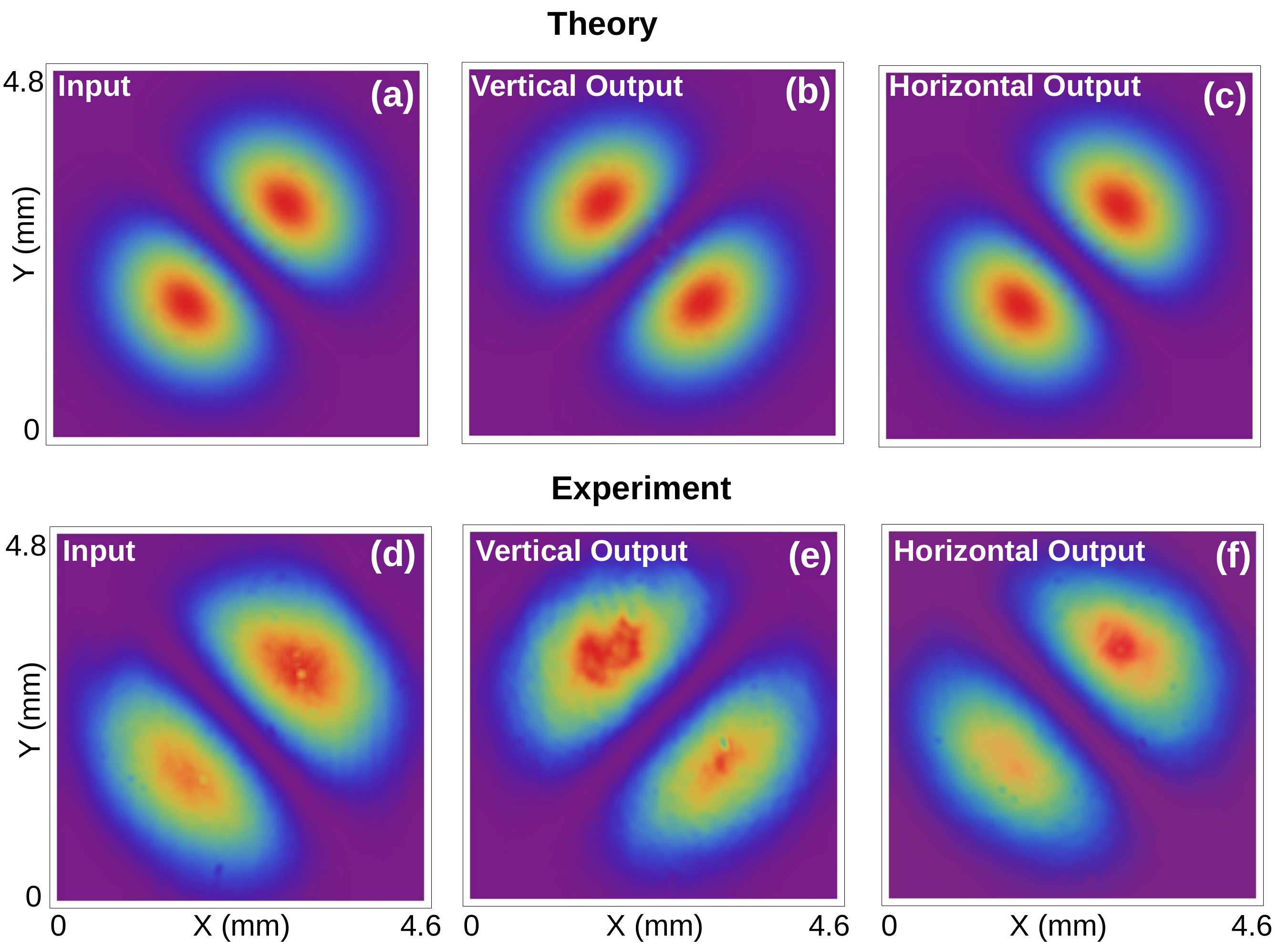}
\caption{Demonstration of the controlled-swap gate. (a) and (d) corresponds to the spatial shape of the input beam. (b) and (e) shows the output beam with vertical polarization where the effect of swap operation can be observed by the change of orientation of the beam with respect to the input beam. (c) and (f) shows the shape of the output beam with horizontal polarization, the same as the one of the input beam. (a), (b) and (c) are theory, (d), (e) and (f) are experimental results.}
\label{fig-dem-swap}
\end{figure}

The c-SWAP gate is a Mach-Zehnder (MZ) interferometer, where light in each arm of the interferometer shows a different polarization. Prior to entering the MZ interferometer, the first Hadamard operation transforms the polarization of the incoming beam into diagonal with the help of a half-wave plate (HWP$_1$). A polarizing beam splitter (PBS$_1$) splits the input beam into the reflected and transmitted beams that have orthogonal polarizations and experience a dissimilar number of reflections given by the number of mirrors present. The OAM of the beams is reversed for an uneven number of reflections and remains the same for an even number of reflections. The phase difference between the two arms is controlled by the displacement of mirrors M$_3$ and M$_4$. 

To verify that the polarization-controlled SWAP gate functions correctly we measure the transverse intensity of the beams with a CCD camera (1200 $\times$ 1600 pixels of $4.4 \times 4.4$~$\mu$m$^2$) before PBS$_1$ (input beam) and after PBS$_2$ (output beam). We image the beams with a telescope with two lenses of focal length 12.5~cm ($L_3$ and $L_4$) and separated by $25$~cm. The CCD is taken away after recording the spatial shape of the beams. We make measurements for light with horizontal and vertical polarizations. 

We use LG modes with index $m=\pm 1$, where the amplitude of the input beam is $\mathrm{LG}_{1}^{0}({\bf r}_{\perp}) + i \mathrm{LG}_{-1}^{0}({\bf r}_{\perp})$, and the polarization is diagonal. Figure \ref{fig-dem-swap} shows the theoretical prediction and the experimental results. The intensity of the input beam is $\sim \rho^2 \exp(-2\rho^2/w_0^2) \cos^2(\varphi-\pi/4)$, where  $\rho$ and $\varphi$ are the radial and azimuthal coordinates, respectively, in cylindrical coordinates and $w_0$ is the beam waist. Figs. \ref{fig-dem-swap}(a) and \ref{fig-dem-swap}(d) show the spatial shape of the input beam, the same for both polarizations. There is a line of zero intensity along $\varphi=3\pi/4$ and $\varphi=-\pi/4$. 

Figures \ref{fig-dem-swap}(b) and \ref{fig-dem-swap}(c) (theory) and Figs. \ref{fig-dem-swap}(e) and \ref{fig-dem-swap}(f) (experiment) show the spatial shape of the output beams. The spatial shape of the beam with horizontal polarization remains unchanged showing the same orientation as the input beam. However, the intensity of the output beam with vertical polarization is $\sim \rho^2 \exp(-2\rho^2/w_0^2) \cos^2(\varphi+\pi/4)]$. It shows zero intensity along the line $\varphi=-3\pi/4$ and $\varphi=\pi/4$, a signature of the effect of the SWAP operation $m \Longleftrightarrow -m$.

The half-wave plate HWP$_2$ performs the second Hadamard operation before detection. Finally, polarizing beam splitter PBS$_3$ separates the horizontal and vertical components of the output beam whose powers ($P_{x}$ and $P_{y}$, respectively) are measured with photodiodes D$_1$ and D$_2$.  

\begin{figure}[t!]
\centering
\includegraphics[width=8.5cm]{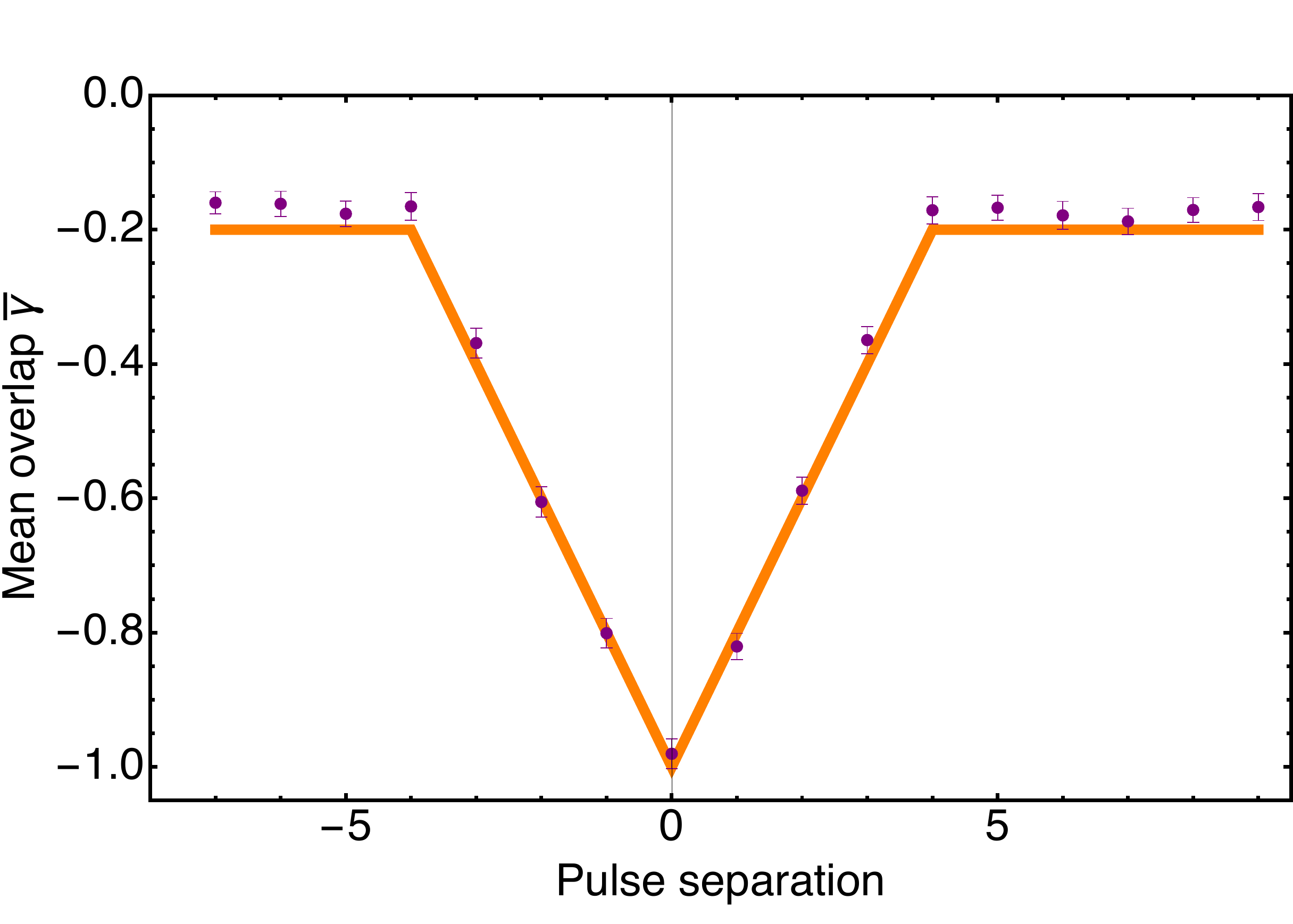}
\caption{Mean overlap $\bar{\gamma}$ between two similar square pulses but delayed one with respect the other. The value of $\bar{\gamma}$ is shown as a function of the pulse separation (in number of times slots $\Delta t$). $\bar{\gamma}=-1$ corresponds to the case when the pulses are not delayed. Dots: experimental data. Solid line: theoretical prediction. See Appendix~\ref{sec: appendix_data} for further details. Error bars represent standard deviation of the value $\bar{\gamma}$.}
\label{fig-square}
\end{figure}

\section{Examples of waveform and data comparison}

Our system allows to compare waveforms and streams of data that vary in time  without measuring its content.  In a series of experiments, we will consider the case that the variables $\{A_m(t_i)\}$ and $\{B_m(t_i)\}$ ($i=1,\ldots,Z$) can take only one of two values: $A_m(t_i),~B_m(t_i)=\pm 1$. This corresponds to encoding logical bits of information "0" and "1" as phases $0$ and $\pi$. The use of this information encoding generates bits in the single mode case ($N=1$) and it generates quarts in the two-modes case ($N=2$). In general there will be $M$ bits (or quarts) whose value will be different, and $Z-M$ bits (quarts) with the same value.  We  define the mean overlap as 
\begin{equation}
\bar{\gamma}=\frac{1}{Z}\,\sum_{i=1}^{Z} \gamma(t_i).
\end{equation}
The mean overlap $\bar{\gamma}$  can be used to estimate how many terms between strings $\left\{ A_m(t_i) \right\}$ and $\left\{ B_m(t_i) \right\}$ are different.
If the two waveforms or strings of data to be compared are equal, one has $\bar{\gamma}=-1$.

A first example is shown in Fig.~\ref{fig-square}, where we measure the mean overlap $\bar{\gamma}$ between two equal square pulses but delayed between them (for further details see Appendix~\ref{sec: appendix_data}). When the two pulses coincide (zero pulse separation) one obtains $\bar{\gamma}= -1$ as expected.

\begin{figure}[t!]
\centering
\includegraphics[width=8.5cm]{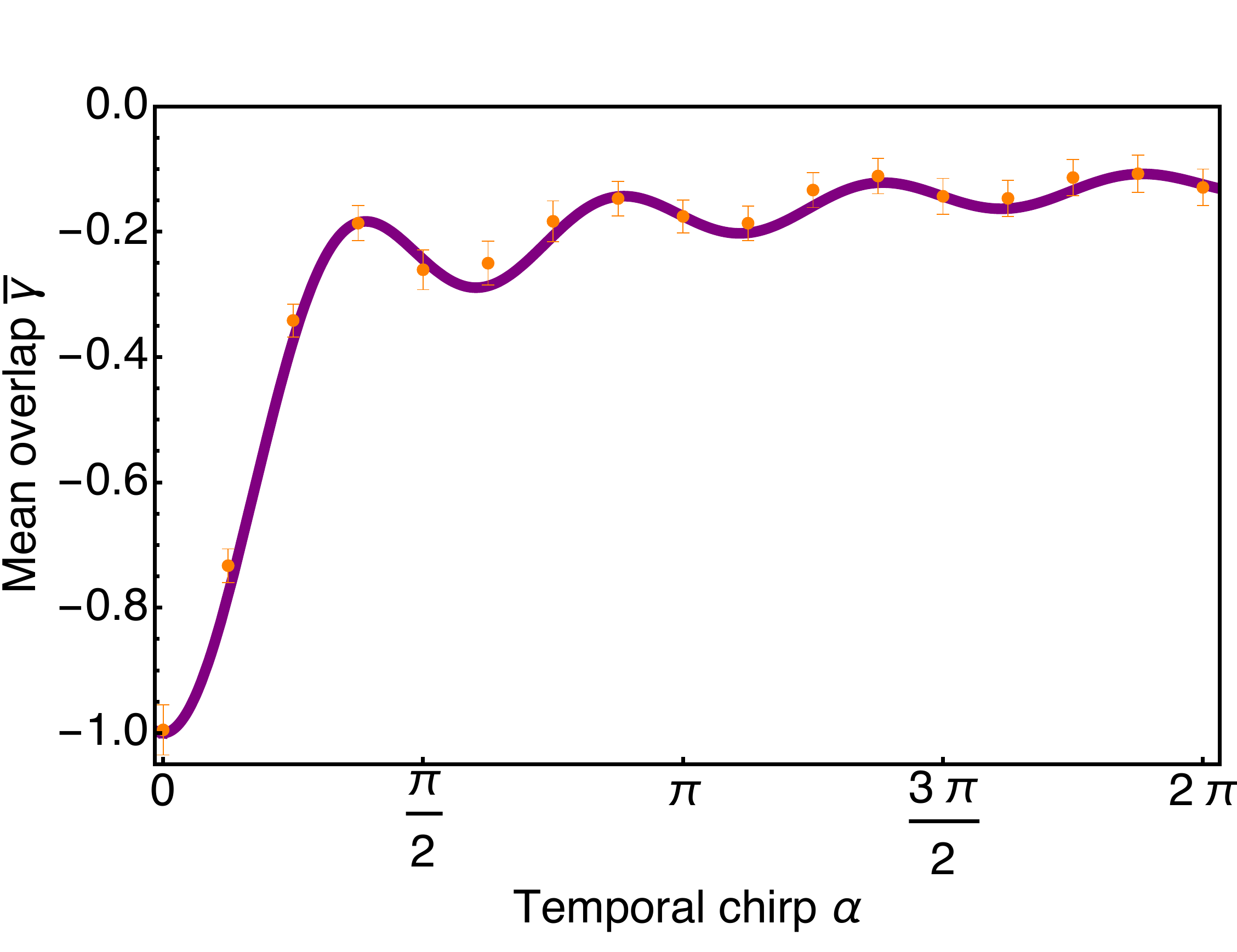}
\caption{Comparison of two signals with different chirp. Signal $A_1$ is constant and signal $B_1$ shows a temporal chirp. Dots: experimental data. Solid line: theoretical prediction. See Appendix~\ref{sec: appendix_data} for further details. Error bars represent standard deviation.}
\label{fig-phase}
\end{figure}

The second example of waveform comparison is shown in Fig.~\ref{fig-phase}. A signal $A_1$ with constant phase is compared with another signal with a chirp $B(t_k)=\exp (i \alpha t_k^2)$ (see Appendix~\ref{sec: appendix_data} for details). $\bar{\gamma}= -1$ corresponds to the case where both waveforms are equal. Increasing the value of the chirp $\alpha$ makes both signals more and more different.   

\begin{figure}[t!]
\centering
\includegraphics[width=8.5cm]{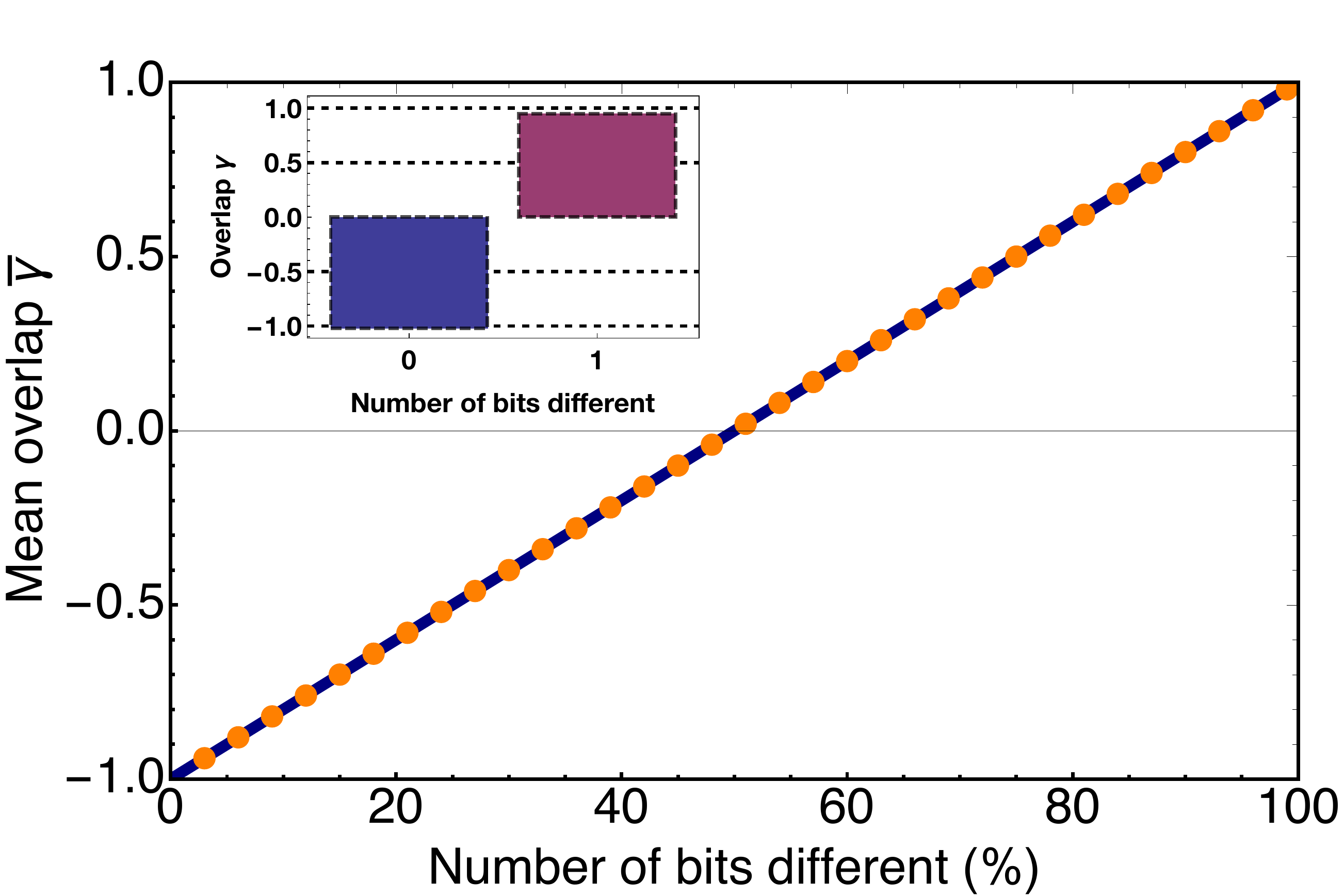}
\caption{
Mean overlap $\bar{\gamma}$ as a function of the fraction of pairs of bits that are different. The solid line corresponds to the expression $\bar{\gamma}=2\,M/Z-1$ (see Appedix~\ref{sec: appendix_data}) where $M$ is the number of pairs $[A_1(t_i),B_1(t_i)]$ where each bit have a different value ($A_1(t_i) \times B(t_i)=-1$) and $Z$ is the total number of pairs of bits. The inset was obtained using 400 different random bits. The figure made use of a subset of 100 random bits from the 400 bits considered in the inset.}
\label{fig-wrong_bit_1}
\end{figure}

We can also compare strings of data. Fig. \ref{fig-wrong_bit_1} shows the experimental result of comparing two strings of random bits at times $t_k$, $A_1(t_k)$ and $B_1(t_k)$, that can take only values of $\pm 1$. $M/Z$ is the fraction of pairs of bits that are different. If the two bits are equal, one obtains $\gamma(t_k)=-1$, while if they are different $\gamma(t_k)=1$. The inset of Fig.~\ref{fig-wrong_bit_1} shows measurements corresponding to the two cases. If the two series of bits are equal ($M=0$), we have $\bar{\gamma}=-1$. If all bits are different ($M=Z$) we have  $\bar{\gamma}=1$. In between, the value of  $\bar{\gamma}$ determines the fraction of bits that are different without the need to evaluate the value of each bit. To correct for the deleterious effect of detection noise in the experiment, we made use of a threshold value to decide when two bits are equal or not: two bits are different if the value measured of  $\gamma(t_k)$ was over 0.7, and they are equal if the value measured was below $-0.7$. 

\begin{figure}[t!]
\centering
\includegraphics[width=9cm]{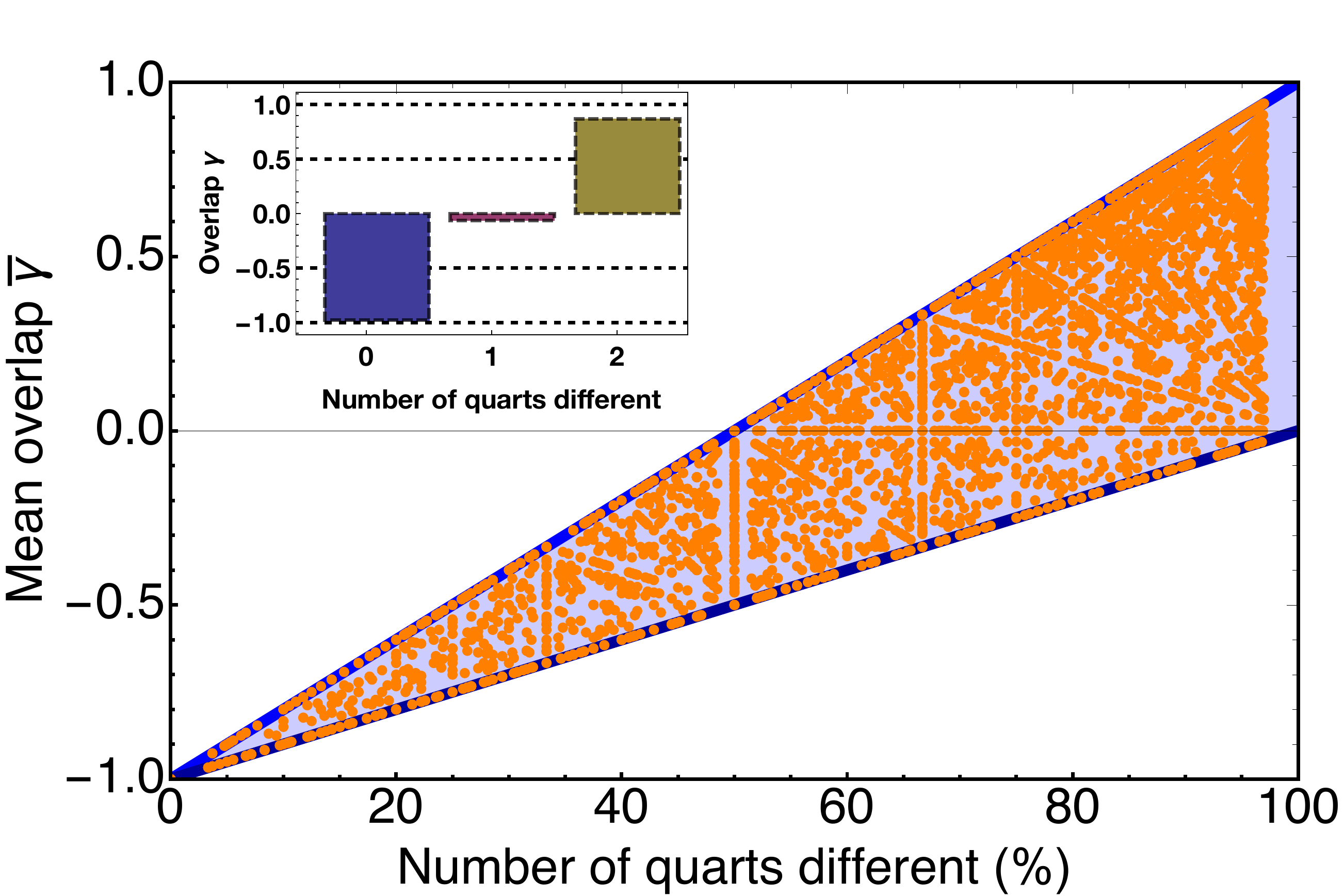}
\caption{Mean overlap $\bar{\gamma}$ as a function of the fraction of pairs of quarts that are different. The inset shows results for the three possible cases:  both quarts are equal ($\bar{\gamma}=-1$), both bits of the quarts are different ($\bar{\gamma}=1$), or just one of the bits of the pair of quarts is different ($\bar{\gamma}=0$). The top solid line $\bar{\gamma}=2\, M/Z-1$ corresponds to the case where all of the errors in a string of quarts are two-bits errors. The lower solid line $\bar{\gamma}=1/2\, M/Z-1$ corresponds to the case where all errors are one-bit errors. The colored region shows the region of possible events. The inset was obtained using 180 quarts, and for the figure, we considered randomly 60 quarts of the previous 180 quarts.}
\label{fig-wrong_bit_2}
\end{figure}

Figure \ref{fig-wrong_bit_2} compares two sets of quarts encoded in the amplitudes of two modes, i.e., [$A_1(t_k)=\pm 1, A_2(t_k)=\pm 1$] and [$B_1(t_k)=\pm 1, B_2(t_k)=\pm 1$]. Differences between quarts can originate from the two bits of the quarts being different ({\em two-bit errors}, $\gamma(t_k)=1$) or just one bit of the quarts being different ({\em one-bit errors}, $\gamma(t_k)=0$). If the two quarts are equal $\gamma(t_k)=-1$. The inset of Fig.~\ref{fig-wrong_bit_2} shows experimental results for all of these possibilities. As shown in the Appendix~\ref{sec: appendix_data}, for a given fraction of different quarts ($M/Z$), the value of $\bar{\gamma}$ ranges between two well-defined values, $M/Z-1$ and $2\, M/Z-1$, corresponding to one- and two-bits errors, respectively. Again, in order to correct for the deleterious effect of detection noise in the experiment, we made use of a threshold value to decide when the bits are equal or not.

\section{Conclusions}
We have demonstrated a protocol for data and waveform comparison that makes use of a Fredkin gate. The functioning of the protocol is a translation of certain features of a counterpart quantum protocol (quantum fingerprinting). The gate uses light beams carrying orbital angular momentum. Intrinsic characteristics of the spatial shape of these modes allow implementing a c-SWAP operation easily, a gate that is generally difficult to implement and that, on many occasions, can only work with a certain probability of success. Our results provide a method to estimate how close are two signals by calculating the overlap between them with simple power measurements. Notice that we can do this in spite that we do not measure the information contained in the signals. The proposed system is another example of the advantages of using light beams with a spatial shape (i.e., structured light).


\section*{Appendix A. Calculation of the overlap factor given in Eq.~(5)}
\label{sec: methods}
The input beam is a superposition of $N$ pairs of orthogonal modes $u_m({\bf r}_{\perp})$ and $v_m({\bf r}_{\perp})$, i.e., $\int d{\bf r}_{\perp}\, u_{m_1}^*({\bf r}_{\perp}) u_{m_2}({\bf r}_{\perp})=\delta_{m_1,m_2}$,  $\int d{\bf r}_{\perp}\, v_{m_1}^*({\bf r}_{\perp}) v_{m_2}({\bf r}_{\perp})=\delta_{m_1,m_2}$, and $\int d{\bf r}_{\perp}\, u_{m_1}^*({\bf r}_{\perp}) v_{m_2}({\bf r}_{\perp})=0$. The electric field writes 
\begin{equation}
\begin{aligned}
{\bf E} ({\bf r}_{\perp}) = \sum^{N}_{m = 1} \left[ A_{m} u_m({\bf r}_{\perp}) + B_{m} v_m({\bf r}_{\perp}) \right] {\bf p},
\end{aligned}
\label{equ:superposition_methods}
\end{equation}
with $\mathbf{p}\equiv\{\mathbf{x},\mathbf{y}\}$, where ${\bf x}$ designates horizontal polarization and ${\bf y}$ designates vertical polarization.  Information is encoded into the complex amplitudes $A_m$ and $B_m$. If one considers the case $A_m,B_m=\pm 1$, bits can be encoded with the help of a single mode: $A_1$ and $B_1$. Quarts require the use of two modes: $A_1,A_2$ and $B_1,B_2$.

In our experimental implementation the orthogonal modes are LG beams with topological index $m$ and radial index $p=0$, which read as
\begin{equation}
u_m({\bf r}_{\perp})=C_{m} \left( \frac{\rho}{w_0}\right) ^{|m|} \exp\left( -\frac{\rho^2}{w_0^2} \right) \exp \left(i m \varphi \right),  
\end{equation}
where $m=1,2,\ldots$. Similarly for modes $v_m$ but with $m=-1, -2,\ldots$. $\rho$ and $\varphi$ are the radial and azimuthal coordinates, respectively, in cylindrical coordinates, $w_0$ is the beam waist and $C_{m}$ is a normalization constant so that $\int \rho d\rho\, d\varphi |u_{m}(\rho,\varphi)|^{2}=1$. 

We first perform a Hadamard operation that transforms the input state with polarization ${\bf x}$ to a diagonal state with polarization $({\bf x}+{\bf y})/\sqrt{2}$.  We use the polarization of the modes as control bit. We implement a polarization-controlled SWAP gate followed by a second Hadamard operation:
\begin{eqnarray} 
& & \sum^{N}_{m = 1} \left[ A_{m} u_m({\bf r}_{\perp}) + B_{m} v_m({\bf r}_{\perp}) \right] {\bf x}  \nonumber \\
& & \xrightarrow{\mathrm{Hadamard 1}} \sum^{N}_{m = 1} \left[ A_{m} u_m({\bf r}_{\perp}) + B_{m} v_m({\bf r}_{\perp}) \right] \frac{{\bf x}+{\bf y}}{\sqrt{2}} \nonumber \\  
& &  \xrightarrow{\mathrm{c-SWAP}}  \sum^{N}_{m = 1} \frac{{\bf x}}{\sqrt{2}} \left[ A_{m} u_m({\bf r}_{\perp}) + B_{m} v_m({\bf r}_{\perp}) \right] \nonumber \\
& & + \frac{{\bf y}}{\sqrt{2}} \left[ B_{m} u_m({\bf r}_{\perp}) + A_{m} v_m({\bf r}_{\perp}) \right] \nonumber \\
& & \xrightarrow{\mathrm{Hadamard 2}} \sum^{N}_{m = 1} \frac{{\bf x}}{2} \left\{ \left( A_{m}+B_m \right) \left[ u_m({\bf r}_{\perp}) + v_m({\bf r}_{\perp})\right] \right\} \nonumber \\
& & + \frac{{\bf y}}{2} \left\{ \left( A_m-B_m\right) \left[ u_m({\bf r}_{\perp})-v_m({\bf r}_{\perp})\right] \right\} .
\end{eqnarray}
Each reflection in a mirror performs the transformation of the topological index $m \Longleftrightarrow -m$. Five reflections along the arm of the interferometer with vertical polarization implement the SWAP operation $A_m \Longleftrightarrow B_m$. The beam that propagates along the arm with horizontal polarization suffers an even number of reflections so that the index $m$ keeps its sign. We should notice that it would also be possible to implement a general transformation $m_1 \Longleftrightarrow m_2$ using a spatial light modulator as demonstrated in \cite{wang}.

With the help of a polarizing beam splitter, we measure the output power carried by modes with orthogonal polarizations:
\begin{eqnarray}
& & P_{y} = \frac{\alpha}{2} \sum_{m = 1}^{N} \vert A_{m} - B_{m} \vert^{2},  \nonumber  \\
& & P_{x} = \frac{\alpha}{2} \sum_{m = 1}^{N} \vert A_{m} + B_{m} \vert^{2}, 
    \label{met-output_power}
\end{eqnarray}
where $\alpha$ is a factor that takes into account the efficiency of detectors and losses of the setup.

In order to evaluate the similarity between two strings of complex numbers $A_m$ and $B_m$, without measuring its content directly, we define the degree of overlap $\gamma$ as:
\begin{equation}
\gamma=\frac{P_{y}-P_x}{P_y+P_x}=- \frac{2\sum_{m=1}^N  \Re{A_m B_m^*}}{\sum_{i=m}^N (\vert A_m\vert^2+\vert B_m\vert^2 ) }.
\label{gamma}
\end{equation}

\section*{Appendix B. Theoretical predictions for data in Figs.~4 -- 7}\label{sec: appendix_data}

\subsection*{B.1. Degree of similarity between two square pulses of the same width but delayed one with respect to the other  (Fig.~4)}\label{sec: appendix_square}
To construct the two square pulses, we consider $20$ times slots. The input signal is thus $A_{1}(t_k) u_{1}({\bf r}_{\perp}) + B_{1}(t_k) v_1({\bf r}_{\perp})$. The pulse defined by $A_1$ is fixed:
\begin{eqnarray}
& & A_{1}=1, \hspace{1cm}  8 \leq k \leq 11, \nonumber \\
& & A_{1}=-1, \hspace{1cm}  \text{elsewhere},   
\end{eqnarray}
while we change the position of the pulse defined by $B_1$:
\begin{eqnarray}
& & B_{1}=1, \hspace{1cm}   l \leq k \leq l+3, \nonumber \\
& & B_{1}=-1, \hspace{1cm}  \text{elsewhere}. 
\end{eqnarray}
The height of the square pulses is $2$ and the width is $4$ time slots. The pulse separation goes from $-7 \Delta t$ to $9 \Delta t$, the shortest distance in time slots between bits with the same value $+1$. When the two square pulses coincide in time ($l=8$) we should measure $\bar{\gamma}= -1$.


\subsection*{B.2. Degree of similarity between two signal with different chirp (Fig.~5)}\label{sec: appendix_chirp}
The input signal is $A_{1}(t_k) u_{1}({\bf r}_{\perp}) + B_{1}(t_k) v_1({\bf r}_{\perp})$. The signal $A(t_k)=1$ is compared with a signal with chirp $B=\exp(i \alpha t_k^2)$. We consider values of $\alpha$ that go from $0$ with $2\pi$ in $16$ steps.

The output power in both orthogonal polarizations are:
\begin{eqnarray}
& & P_x=\alpha \left[1+\cos(\alpha t_k^2) \right], \nonumber \\
& & P_y=\alpha \left[1-\cos(\alpha t_k^2) \right].
\end{eqnarray}
The overlap $\gamma_k$ at time slots $k \Delta t$ is 
\begin{equation}
    \gamma_k = -\cos{\alpha t_k^{2}}. 
\end{equation}
We measure the sum of all overlap $\gamma_k \tau$ where we choose $\tau=T/40$. When we substitute the sum for an integral we obtain 
\begin{equation}
    \bar{\gamma} = -\frac{1}{T}\int_0^T \cos(\alpha \tau^2) d\tau= - \sqrt{\frac{\pi}{2 \alpha}}\mathrm{FresnelC}\left[ T \sqrt{\dfrac{2 \alpha}{\pi}} \right],
\end{equation} 
where $\mathrm{FresnelC}(X)$ is the so-called Fresnel cosine function. For no chirp ($\alpha=0$) the two waveforms are equal and one has $\bar{\gamma} = -1$.


\subsection*{B.3. Comparison between two strings of bits (Fig.~6)}\label{sec: appendix_string}
The amplitude of the electric field writes
\begin{equation}
{\bf E} ({\bf r}_{\perp}) = A_1 u_1({\bf r}_{\perp}) + B_1 v_1({\bf r}_{\perp}).
\end{equation}
From Eq. (\ref{gamma}) one obtains that the overlap is $\gamma=-A_1 B_1$. If the two bits are equal, we have $\gamma=-1$, if they are different we have $\gamma=1$. The variable $\bar{\gamma}$ is
\begin{equation}
{\bar \gamma}=2\frac{M}{Z}-1,
\label{calculation_figure6}
\end{equation}
where $M/Z$ is the fraction of pairs of bits with a different value.


\subsection*{B.4. Comparison between two strings of quarts (Fig.~7)}
The quart is encoded in the amplitudes of two modes. The amplitude of the electric field writes
\begin{equation}
\begin{aligned}
{\bf E} ({\bf r}_{\perp}) = A_{1} u_1({\bf r}_{\perp}) +A_{2} u_2({\bf r}_{\perp})+ B_{1} v_1({\bf r}_{\perp}) + B_{2} v_2({\bf r}_{\perp}).
\end{aligned}
\end{equation}
The overlap is
\begin{equation}
    \gamma=-\frac{A_1 B_1+A_2 B_2}{2}.
\end{equation}
There are three possibilities:
\begin{itemize}
    \item The two quarts $A_{1,2}$ and $B_{1,2}$ have the same value: $A_1=B_1$ and $A_2=B_2$. The overlap is $\gamma=-1$.
    \item The two bits of the quart are different: $A_1 \neq B_1$ and $A_2 \neq B_2$. The overlap is $\gamma=1$.
    \item A pair of bits of the quart are different: $A_1 \neq B_1$ or $A_2 \ne B_2$, but the remaining bit is equal. The overlap is $\gamma=0$.
\end{itemize}
$M$ pairs of quarts encode a different value, maybe because both bits of the quart are different or because one of the bits are different. In any case this makes the quarts to be different. The variable $\bar{\gamma}=1/Z\,\sum_k \gamma_k$ can take a range of values that depends on the fraction of pairs of quarts that encode different information ($M/Z$). If all difference between quarts are one-bit differences
\begin{equation}
{\bar \gamma}_1=\frac{M}{Z}-1.
\label{calculation_figure71a}
\end{equation}
If all difference are two-bits differences
\begin{equation}
{\bar \gamma}_2=2\frac{M}{Z}-1.
\label{calculation_figure71b}
\end{equation}
In general, for two arbitrary strings of quarts encoded in the way described above, $\bar{\gamma}$ will have a value larger than $\gamma_1$ but lower than $\gamma_2$.


\section*{Appendix C. Influence on experimental data of noise detected by non-ideal detectors}
How the value measured of the overlap $\gamma$ changes when one considers the signal detected (background noise) of non-ideal detectors? For the sake of simplicity, let us consider the case where we compare two strings of bits encoded in a single mode. 

When we measure experimentally the power in the vertical and horizontal polarizations, we will obtain
\begin{eqnarray}
& & P_{y} = I_{0} + C, \nonumber \\
& & P_{x} = C_0,
\end{eqnarray}
for different bits, and 
\begin{eqnarray}
& & P_{y} =  C_0, \nonumber \\
& & P_{x} = I_0+C_0,
\end{eqnarray}
for equal bits. $I_0$ would be the total power detected with ideal detectors and $C_0$ is the background noise measured when no input is considered. When varying the degree of difference between bits, we can measure the visibility as
\begin{equation}
\text{V}=\frac{P_{y,max}-P_{y,min}}{P_{y,max}+P_{y,min}}=\frac{I_0}{I_0+2C_0}.
\end{equation}
The experimentally measured value $\bar{\gamma}_{exp}$ compared with the ideal value $\bar{\gamma}_{ideal}$  that would be obtained with ideal detectors is
\begin{equation}
\gamma_{\mathrm{exp}} =\frac{1}{N} \left[  \frac{M\, I_0}{ I_0+2C_0}-\frac{(N-M)\, I_0}{I_0+2C_0} \right]=
\text{V}\,\gamma_{\mathrm{ideal}}. \nonumber
\end{equation}

\section*{Acknowledgments}
We acknowledge financial support from the Spanish Ministry of Economy and Competitiveness through the ``Severo Ochoa'' program for Centres of Excellence in R\&D (SEV-2015-0522), from Fundaci\'{o} Privada Cellex, from Fundaci\'{o} Mir-Puig, and from Generalitat de Catalunya through the CERCA program.

\section*{Funding}
Spanish Ministry of Economy and Competitiveness, “Severo Ochoa” Programme for Centres of Excellence in R$\&$D (SEV-2015-0522, PRE2018-085072). Consejo Nacional de Ciencia y Tecnolog\'{i}a, CONACYT (257517, 280181, 293471, APN2016-3140). Secretar\'ia de Educaci\'on, Ciencia, Tecnolog\'ia e Innovaci\'on, SECTEI (SECITI/073/2017, SECTEI/178/2019).


\section*{Disclosures}
The authors declare no conflicts of interest.

\bibliography{references}

\end{document}